\documentclass[conference]{IEEEtran}
\usepackage{xspace}
\usepackage{amsfonts, amsmath, amssymb}
\usepackage{graphicx}
\usepackage{hyperref}
\usepackage{cleveref}
\usepackage{xcolor}

\def\name{{\sc PIEChain}\xspace}

\begin{document}
\title{Demo: \name ~- A Practical Blockchain Interoperability Framework}

\author{
\IEEEauthorblockN{Dani\"el Reijsbergen,$^*$ Aung Maw,$^\dagger$ Jingchi Zhang,$^*$ Tien Tuan Anh Dinh,$^\ddagger$ and Anwitaman Datta$^*$ \\[0.2cm]}
\IEEEauthorblockA{$^*$Nanyang Technological University, Singapore, Singapore \\
$^\dagger$Singapore University of Technology and Design, Singapore, Singapore \\
$^\ddagger$Deakin University, Melbourne, Australia\vspace{-0.1cm}}

}

\maketitle

\begin{abstract}
A plethora of different blockchain platforms have emerged in recent years, but many of them operate in silos. As such, there is  a need for reliable \textit{cross-chain} communication to enable blockchain interoperability. Blockchain interoperability is challenging because transactions can typically not be reverted -- as such, if one transaction is committed then the protocol must ensure that all related transactions are committed as well. Existing  interoperability approaches, e.g., Cosmos and Polkadot, are limited in the sense that they only support interoperability between their own subchains, or require intrusive changes to existing blockchains. To overcome this limitation, we propose \name, a general, Kafka-based cross-chain communication framework. 
We utilize \name for a practical case study: a cross-chain \textit{auction} in which users who hold tokens on multiple chains bid for a ticket sold on another chain. \name is the first publicly available, practical implementation of a general framework for cross-chain communication.
\let\thefootnote\relax\footnotetext{{This work was supported by Ministry of Education (MOE) Singapore's Tier 2 Grant Award No. MOE-T2EP20120-0003.}}
\end{abstract}




\section{Introduction}

A blockchain is a replicated, tamper-evident database designed for hostile environments. It consists of a
number of nodes, of which some may be malicious, which maintain an append-only \textit{ledger}. The ledger stores {\em transactions} that modify some global states. In the canonical example,
i.e., cryptocurrencies~\cite{btc_origin}, the global states are user accounts and native (fungible) tokens,
and the ledger contains transactions transferring tokens from one account to another. In another emerging application, the blockchains store \textit{non-fungible tokens} (NFTs) uniquely representing assets, e.g., digital art works or concert tickets. Many 
blockchains also support \textit{smart contracts} letting users define their own states and codes that modify the
states. Smart contracts are stored in the ledger and are replicated and kept consistent by all
the blockchain nodes. 

\begin{figure}
  \centering
  \includegraphics[width=\linewidth]{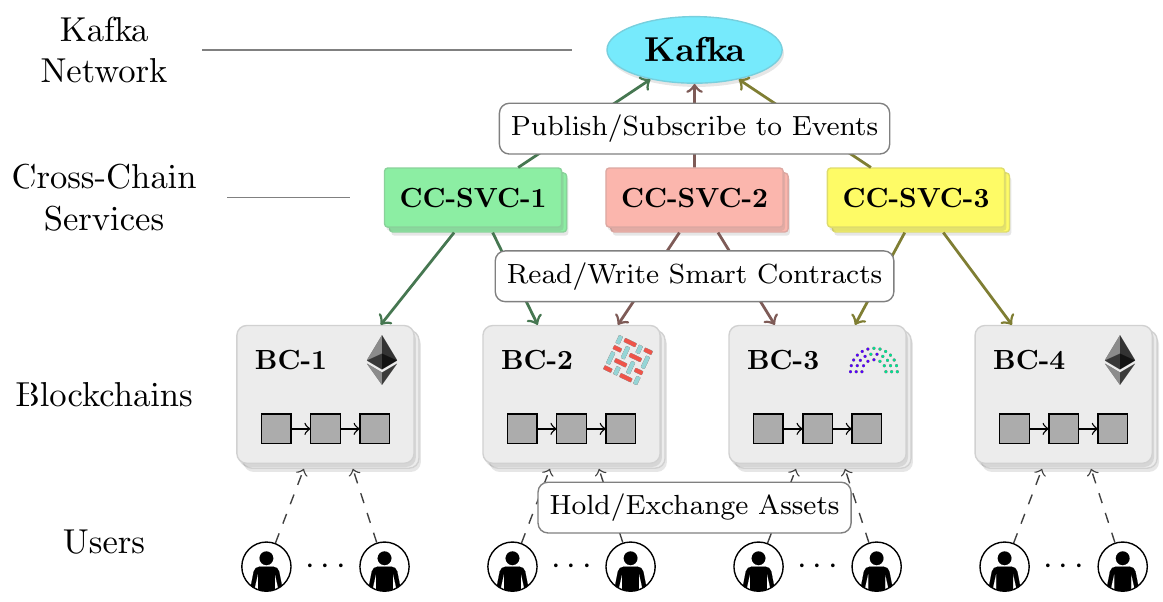}
  \caption{\name architecture: cross-chain services (CC-SVCs) read/write events from/to the Kafka network, and interact with the different underlying blockchains (BCs).}
  \label{fig:ccc_design}
\end{figure}

In recent years, many independent blockchain platforms have emerged, resulting in an ecosystem with a long tail. On the one hand, there are a small number
of hugely popular, general-purpose blockchain platforms such as Ethereum and Hyperledger. On the other hand,
there are thousands of smaller blockchains designed for specific applications, most of which are only in early
stages of development. This includes more than $10\,000$
cryptocurrencies \cite{coinlore} as of early 2023, alongside blockchains for healthcare, identity management, and IoT. In general, these
blockchains do not interoperate, i.e., they exist in silos. As such, blockchain \textit{interoperability}, i.e., the ability for users to exchange information or assets on different blockchains, is a subject that is drawing increasing interest from the research community \cite{herlihy2021cross,zamyatin2021sok,engel2021failure,belchior2021survey,zamyatin2019xclaim}. 

One of the main challenges in designing a secure blockchain interoperability framework is to guarantee \textit{atomicity}, i.e., that either all steps of an agreed set of transactions terminate successfully, or none do. This is more complicated in blockchains than in traditional databases because blockchain transactions are (in principle) irreversible. For example, if a payment for an NFT on Chain A has already been made on Chain B, then atomicity requires that the transaction on Chain A \textit{must} go ahead because the transaction on Chain B cannot be reverted. One common approach to guarantee atomicity is to \textit{escrow} all of the tokens that are involved in the deal in smart contracts, and release them only through a \textit{commit} message that is signed by all parties \cite{herlihy2021cross}. Another important challenge in blockchain interoperability is to guarantee \textit{liveness}, so that the tokens that are escrowed cannot be frozen forever. To ensure liveness, it must be possible for nodes to send an \textit{abort} message that allows all parties to withdraw their assets from escrow. 

To guarantee both atomicity and liveness, an interoperability framework must be able to tolerate \textit{adversarial} nodes who send a commit message to one blockchain and an abort message to another.
It is known for asynchronous networks (i.e., where there is no bound on message delays) that this is impossible to achieve  without a trusted third party (TTP) \cite{zamyatin2021sok}. There are two main approaches to overcome this challenge \cite{herlihy2021cross}. The first approach is to combine a synchrony assumption with a cooldown period for aborts -- i.e., assuming that a commit vote, once generated, can be appended to all affected blockchains before the end of any cooldown period for aborts. This approach is taken in, e.g., hashed timelocked contracts \cite{herlihy2018atomic}. The second approach is to replace the TTP with another blockchain to ensure that either a valid commit or abort message can be created, but not both \cite{herlihy2021cross,zakhary2021atomic}. 

Although approaches of both types have been proposed in the scientific literature, they either do not have a publicly available implementation \cite{herlihy2018atomic, herlihy2021cross,zakhary2021atomic}, or are limited in their application scope, e.g., creating backed assets on another blockchain \cite{zamyatin2019xclaim} or token swaps \cite{thyagarajan2022universal}.
In a separate development, several blockchain platforms have emerged that enable interoperability by default, e.g., Cosmos and Polkadot. However, these platforms only support interoperability between their own subchains, or require intrusive changes to existing blockchains. This motivates the need for a \textit{general} and \textit{practical} interoperability framework that can be interfaced to existing blockchain platforms without modifying them. 
Our goal is to provide such a framework.

\paragraph*{Contributions \& Novelty} We present \name
to achieve this goal. As synchrony is difficult to assume in practice, \name uses \textit{cross-chain services} to replace the TTP. The cross-chain services communicate using an event log that uses Apache's Kafka protocol for efficiency. To demonstrate the practical relevance of \name, we have implemented a cross-chain service for a realistic case study: a cross-chain auction. We have interfaced \name with some of the most popular blockchain platforms that support smart contracts: Ethereum, Hyperledger Fabric, and Quorum. Finally, we have developed a GUI for our case study. {The auction case study is one of three case studies (two auctions and a flash loan \cite{herlihy2021cross}) whose code can be found in the PIEChain code repository on GitHub.\footnote{\url{https://github.com/aungmawjj/piechain}}

\section{Overview of \name}

\subsection{Entities}
\label{sec:entities}

The main entities in \name are as follows (see also \Cref{fig:ccc_design}):

 \textit{Blockchains}, which store \textit{assets} (e.g., tokens, keys) that are held by \textit{users}. A user may hold assets on multiple blockchains. Each blockchain has its own protocol to determine who has read and write access  --  blockchains are typically either \textit{permissioned}, i.e., a fixed set of nodes has read and write access, or \textit{permissionless}, i.e., everyone has read access and can create transactions, and nodes with sufficient power (e.g., processing speed) can add transactions to the blockchain.
 
 \textit{Cross-chain services} (CC-SVCs), which allow users to exchange assets on different blockchains. Each CC-SVC consists of a server that interacts with user clients to facilitate cross-chain communication. In practice, the CC-SVC charges the users for participation, and it can interact with any number of blockchains. In the following, each CC-SVC corresponds to a set of events involved in an atomic exchange of assets that are sent by a server to the Kafka ledger. In practice, a single server may operate many CC-SVCs.
 
The \textit{Kafka network}, which serves as an append-only log of events generated by the CC-SVCs.  Events correspond to transactions made on underlying blockchains. The Kafka network is operated by a fixed set of nodes who charge CC-SVCs for uploading events.

In \name, we assume that the CC-SVCs are semi-trusted, and that they are motivated to behave honestly by having a reputation to uphold, akin to commercial outsourced service providers. The Kafka network operators are untrusted but have no incentive to misbehave as they do not interact with the underlying blockchains. 
This allows us to run a protocol (Kafka) that prioritizes efficiency over security. 
An alternative design is to make the CC-SVCs untrusted and the Kafka network trusted. In this case, each Kafka node runs a (light) client for each of the underlying blockchains to validate the inclusion of transactions on those chains. In this case, the event log would have to use a more secure protocol such as PBFT \cite{castro1999practical}. We leave such a design as future work.

\subsection{Process Flow}
\label{sec:process_flow}
 
Given the entities of \Cref{sec:entities}, the process flow in \name has the same structure as \textit{cross-chain deals} as proposed by Herlihy et al.\ \cite{herlihy2021cross}. Cross-chain deals are agreements by multiple users to exchange assets on different blockchains, and consist of five phases (see also \Cref{fig:steps}):
\begin{enumerate}
\item \textit{Clearing phase}: the CC-SVC creates the smart contracts on the different blockchains that are used to escrow and transfer the assets involved in the deal.
\item \textit{Escrow phase}: the users escrow their outgoing assets by transferring them to the smart contracts.
\item \textit{Transfer phase}: the assets are tentatively exchanged, i.e., the execution logic of the smart contracts is specified. 
\item \textit{Validation phase}: each user checks whether the result of the execution logic would be to their satisfaction.
\item \textit{Commit phase}: the deal is concluded through \textit{commitment} if all parties are satisfied, or through \textit{abortion} otherwise. Commitment means that the execution logic in the smart contracts is executed and that the assets are exchanged as specified by the deal. Abortion means that the assets in each smart contract are returned to their original owners.
\end{enumerate}
To commit, the users interactively construct a \textit{commit vote}, which is sent by the CC-SVC to the Kafka ledger. To abort, a single user sends an abort message to the CC-SVC. For each CC-SVC, either a commit or abort message can be added to the Kafka ledger, but not both. An inclusion proof of a commit vote on the Kafka ledger is accepted by all the smart contracts on the different blockchains -- this guarantees that once a commit vote is constructed, either all asset transfers can be executed or none. 

\begin{figure}
  \centering
  \includegraphics[width=\linewidth]{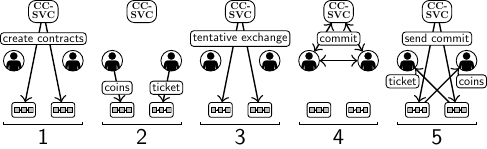}
  \caption{Illustration of the five steps of \Cref{sec:process_flow} in a setting with one CC-SVC (top), two users (middle), and two blockchains (bottom). The Kafka network is not shown.}
  \label{fig:steps}
\end{figure}

\section{Implementation of \name}

To illustrate the practical applicability of \name, we have interfaced it to several commonly-used blockchain platforms and used it to implement an application from the scientific literature \cite{herlihy2021cross}: a cross-chain auction for a digital asset. The blockchain support extends to other case studies, as we discuss in \Cref{sec:extensions}.

\subsection{Blockchain Support}

To interface an underlying blockchain with \name, the CC-SVCs must be able to validate transactions on those chains. 
{Our implementation supports the following blockchain platforms: \textit{Ethereum} (both the Proof-of-Work and Proof-of-Authority versions of private Ethereum), \textit{Hyperledger Fabric}, and \textit{Quorum}. The latter two support permissioned blockchains, whereas Ethereum has a permissionless main chain but also supports private chains with the same functionality.}

\subsection{Auction}

In our case study, an auctioneer sells an asset on one blockchain and receives payment in the form of assets on another blockchain. As in \cite{herlihy2021cross}, we use the example of a ticket seller. The ticket is an NFT on a dedicated blockchain, whereas the other blockchain supports more commonly-used fungible tokens (e.g., Ether). The former blockchain is called the \textit{ticket blockchain} and the latter the \textit{coin blockchain}. This is easily generalized to settings with more than one coin blockchain. In the following, we consider a first-price auction (i.e., the bidder with the highest bid pays its bid and receives the asset). The five phases of the protocol are then as follows:

\begin{enumerate}
\item The auctioneer enlists a CC-SVC which creates smart contracts on the ticket blockchain and the coin blockchains.

\item The auctioneer transfers its asset (the ticket's NFT) to the ticket contract, and the bidders transfer their bids to the contract on their coin blockchain.

\item The execution logic is determined: the auctioneer updates each of the ticket and coin contracts by specifying which party receives the ticket and which bid is transferred to the auctioneer. (Note that this logic cannot be specified in the ticket contract \textit{a priori} because the contract on the ticket chain cannot read data from the coin blockchains.)

\item Each user (i.e., the auctioneer and the bidders) determines whether the outcome of the transfer protocol is agreeable to them, i.e., whether the ticket is indeed transferred to the correct party.

\item All users construct a commit vote -- once this has been constructed, it is sent to each contract to enact the transfers specified in the transfer phase.

\end{enumerate}
In \name, the auction requires two (logical) types of CC-SVC: the \textit{relayer} and the \textit{signer}. The relayer listens for events (bids) on the coin chains and relays them to the ticket blockchain. The \textit{signer} assists in creating the commit vote. 

\section{Demonstration Plan}

For our demonstration, we have developed a graphical user interface (GUI) using the React framework to illustrate a cross-chain auction. 
The GUI consists of three main pages: a dashboard page that displays a list of known auctions as depicted in \Cref{fig:dashboard}, a detailed view for individual auctions as depicted in \Cref{fig:auction_view}, and a page for the creation of new auctions (not displayed). The view is the same for an auctioneer as for bidders. On the dashboard view, prospective auctioneers can click the ``Create New Auction'' button to initiate an auction -- the auctioneer selects a CC-SVC, the asset to be auctioned, which other blockchains to accept bids from, the exchange rate for tokens between different blockchains (which must be fixed in advance), and the time at which the auction is concluded. Next, the relay CC-SVC creates the relevant contracts, and sends the address of the contract on the asset chain to the auctioneer. The auctioneer can then add the auction on its dashboard by adding the contract address and clicking the ``Add Existing Auction'' button. Meanwhile, it advertises the contract address to potential bidders. 

\begin{figure}
  \centering
  \includegraphics[width=0.8\linewidth]{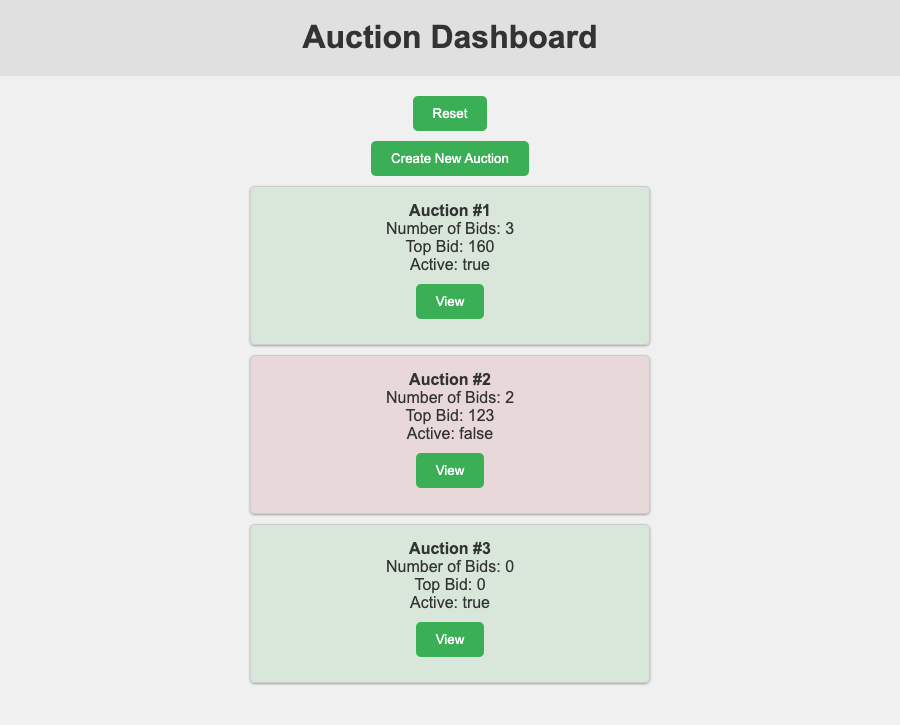}
  \caption{Auction dashboard.}
  \label{fig:dashboard}
\end{figure}

When the bidders become aware of the asset contract address, they can also add it to their dashboards. After the bidder has added the auction, she can view it in more detail by pressing the ``View'' button in the auction's panel, which takes her to the detailed view page. On this page, the bidder is able to view crucial information about the auction such as its creation and conclusion times, and a list of bids. The highest bid is marked with an asterisk. If the auction is still ongoing, then the bidder can add a bid by specifying the blockchain on which the bid is made and the amount of tokens to bid. A transaction is then made to the relevant coin contract and the information is sent to the relay CC-SVC. A user can also abort any bids that she has made via the CC-SVC.

After the auction is concluded, the relay CC-SVC informs all participants and specifies the tentative transfer of goods in the smart contracts. The CC-SVC then asks all participants' clients to participate in creating a commit vote. (Failure to participate should result in a penalty \cite{engel2021failure}.) When the commit vote has been created, it is sent to all contracts to initiate the final transfer of assets. At this point, the GUI will show that the auction has concluded.

\begin{figure}
  \centering
  \includegraphics[width=\linewidth]{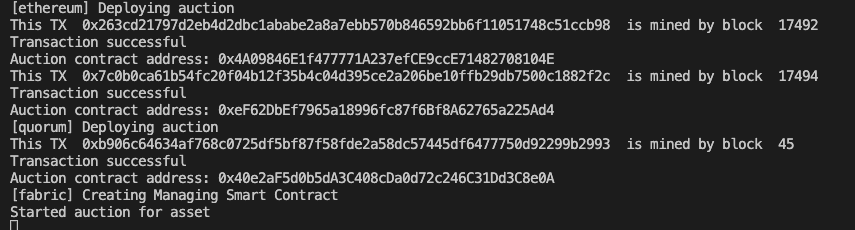}
  \caption{Window showing the console output of a blockchain client.}
  \label{fig:terminal}
\end{figure}

The exact flow of the demo will be as follows:

\begin{enumerate}

    \item One user -- the auctioneer -- opens a web-browser based GUI and uses it to initiate an auction for a selected asset. In this process, all the features on the auction creation page are demonstrated. The asset exists on a dedicated ticket chain running in Hyperledger Fabric. Contracts are created on all involved blockchains {(step 1 of \Cref{fig:steps}}.

    \item At least two other users, who use browser windows on different machines, navigate to the newly created auction's detail page and submit their individual bids for the asset {(step 2 of \Cref{fig:steps})}. At least one bidder uses (private) Ethereum, and the other  uses Quorum. 
		
    \item After some time, the auction is concluded and the winning bid is determined {(step 3 of \Cref{fig:steps})}. This will cause an ``end auction'' event to be sent to the relay CC-SVC by the auctioneer. The signers, who are listening for this type of event, will notice the event and construct a commit vote {(step 4 of \Cref{fig:steps})}. The commit vote is then sent to Kafka, and forwarded by the relay node to the auction contract and the coin chain contracts. At this point, the asset is transferred to the winning bidder, and the winning bid to the auctioneer {(step 5 of \Cref{fig:steps})}.

\end{enumerate}
Throughout the demo, we will use a terminal window to query the states of the underlying blockchains after each step, as displayed in \Cref{fig:terminal}. This will allow the audience to observe the changes happening in the background, and to interact with the flow of the demonstration: e.g., by asking for new actions to see how the background states change.

{To illustrate the flow of the demo, a video can be found online \footnote{\url{https://youtu.be/SRnlszBVWXo}} which depicts the tentative demo slides, and the screen if the auctioneer and bidders were to perform their actions using a single computer. (This is not a limitation of \name, but makes it easier to record video.)}

\section{Extensions}
\label{sec:extensions}

The CC-SVC framework and the interface for the supported blockchains can be used to easily extend \name to other use-cases. One of these is a cross-chain \textit{flash loan} as described in \cite{herlihy2021cross}. A GUI for flash loans would have limited practical relevance as arbitrage opportunities are typically resolved rapidly, so the interactions with the CC-SVC would normally be done by trading bots. However, if time permits, we will display a visualization for the impact of the steps in a flash loan on the states of the various involved contracts.

\begin{figure}
  \centering
  \includegraphics[width=0.7\linewidth]{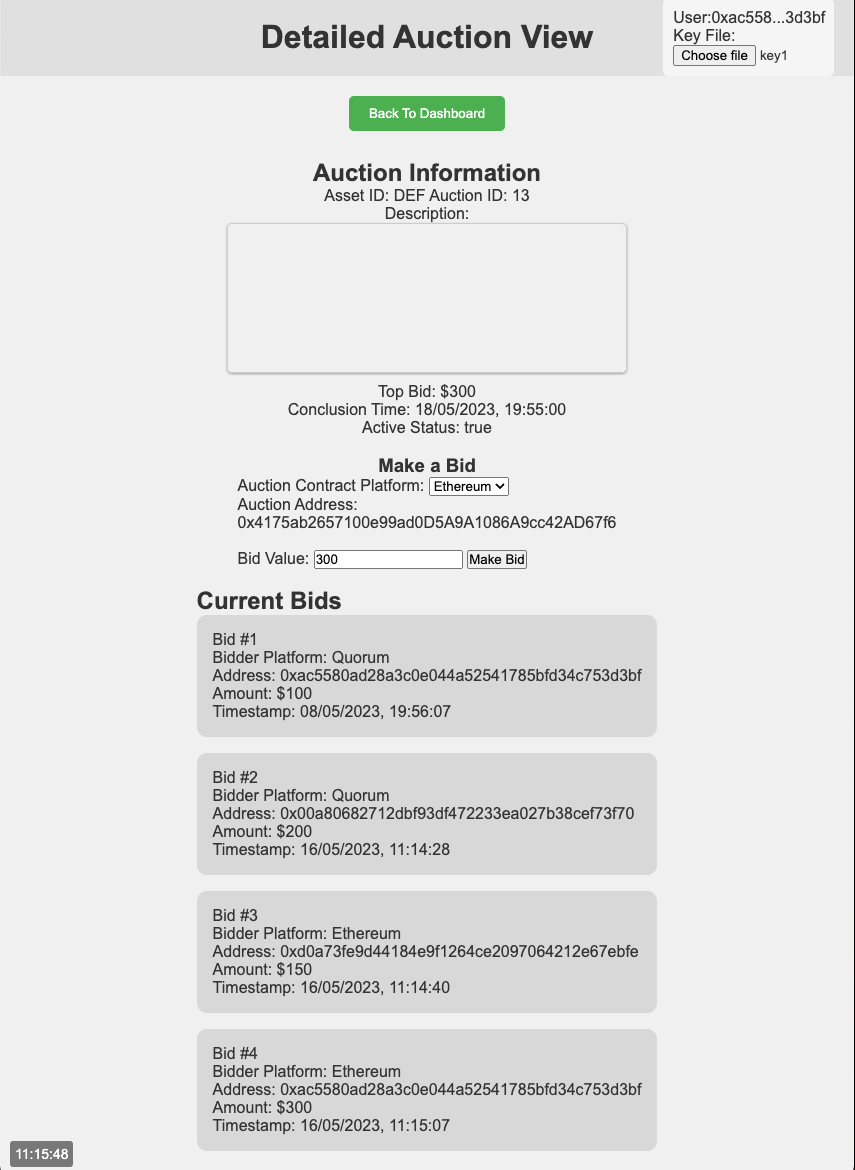}
  \caption{Detailed view of an active auction.}
  \label{fig:auction_view}
\end{figure}

\bibliographystyle{plain}
\bibliography{ref}

\end{document}